\documentclass{emulateapj}
\usepackage{graphicx}
\begin{document}

\title{GAMMA-RAY FLARES AND VLBI
OUTBURSTS OF BLAZARS}
\author{ M.M. Romanova}
\affil{Space Research Institute,
Russian Academy of Sciences, Moscow, Russia; and\\
Department of Astronomy, Cornell University, Ithaca, NY 14853-6801;
romanova@astrosun.tn.cornell.edu}

\author{ R.V.E. Lovelace}

\affil{Department of Astronomy, Cornell University,Ithaca, NY 14853-6801;
rvl1@cornell.edu}

\begin{abstract}
A model is developed for the time
dependent electromagnetic - radio
to gamma-ray - emission of active galactic
nuclei, specifically, the blazars,
based on the acceleration and creation of
leptons at a propagating discontinuity or
{\it front} of a self-collimated
Poynting flux jet. The front
corresponds to a discrete relativistic jet
component as observed with
very-long-baseline-interferometry (VLBI).
Equations are derived for the number, momentum,
and energy of particles in the front taking
into account synchrotron, synchrotron-self-Compton (SSC),
and inverse-Compton processes
as well as photon-photon pair production.
The apparent synchrotron, SSC, and inverse Compton
luminosities as functions of
time are determined. Predictions
of the model are compared with observations
in the gamma, optical, and radio bands. The
delay between the high-energy gamma-ray flare
and the onset of the radio is explained by
self-absorption and/or free-free absorption by
external plasma.  Two types of gamma-ray
flares are predicted, Compton dominated
or SSC dominated, depending on the initial
parameters in the front.
The theory is applied
to the recently observed gamma-ray
flare of the blazar PKS 1622-297
(Mattox et al. 1996).
Approximate agreement of theoretical and
observed light curves is obtained for a viewing
angle $\theta_{obs} \sim 0.1$ rad, a black hole
mass $M\sim 3 \times 10^9 M_\odot$, and a magnetic
field at the base of the jet $B_o \sim 10^3$ G.
\end{abstract}

\section{INTRODUCTION}

New understanding
of the nature of Active Galactic Nuclei
(AGNs) has come from the discovery
of the high energy gamma-ray radiation
in the range $50-10^4$ MeV by the EGRET
instrument on the Compton gamma-ray
Observatory (e.g., Hartman et al. 1992;
Mattox et al. 1993; Thompson et al. 1993). This radiation
is observed from a sub-class of AGNs
termed blazars, which include
Optically Violently Variable (OVV)
quasars and BL Lac objects
and which show strong variability
in all wavebands from
radio to gamma. Many of the objects
reveal `super-luminal' jets in
VLBI maps which indicate that we
observe matter of the jet pointed
nearly towards us and that the
jet matter moves
with relativistic speed.

Prior to the Compton Observatory measurements,
prediction of strong, collimated
gamma-ray emission from AGN relativistic
jet sources was made by the electromagnetic
cascade model of Lovelace, MacAuslan, and
Burns (1979) and Burns and Lovelace (1982).
More recently, a number of theoretical
models have been developed to explain
the observed gamma-ray emission of AGNs (see review
by Sikora 1994). In most of the models the gamma-ray
radiation is ascribed to inverse-Compton
scattering of relativisitic electrons
and possibly positrons (Lorentz factors
$\gamma \sim 10^2-10^3$) of a jet having
relativistic bulk motion (Lorentz factor $\Gamma \sim 10$)
with soft photons (energies $\sim 1-10^2$ eV). The soft photons can
arise from the synchrotron emission of the relativistic
electrons in the jet as in the synchrotron-self-Compton
(SSC) models (Maraschi, Ghisellini,
and Celotti 1992; Marscher and Bloom, 1992)
or from the direct
or scattered thermal radiation from an
accretion disk (Dermer, Schlickeiser, and
Mastichiadis 1992; Blandford 1993; and Sikora,
Begelman, and Rees 1994),
or from a single cloud (Ghisellini and Madau 1996).
The spectrum of blazars in a pair cascade model was
calculated by Levinson and Blandford (1995) and
Levinson (1996).
In different models, ultra high energy protons
(Lorentz factors $>10^6$) are postulated to
cause a cascade, the product particles of which
produce the observed radiation (Mannheim and Biermann 1992),
or the jets considered to be ultra relativistic with
bulk Lorentz factors $\Gamma >10^4$ (Coppi, Kartje, and
K\"onigl 1993).

Here, we propose that the main driving
force for the observed superluminal jet
components is a finite amplitude discontinuity
in a Poynting flux jet.  A brief discussion of the
model was given earlier (Lovelace and Romanova 1996).
A rapid change in the Poynting jet outflow from a
disk can result from implosive accretion in a disk
with an ordered magnetic field (Lovelace, Romanova, and Newman 1994).
Propagation of
newly expelled EM field and matter
from the disk with higher velocity than
the old jet
can under certain conditions lead to
the formation of a pair of shock
waves as in the case of non-relativistic
hydrodynamic flow (Raga et al. 1990).
Particle acceleration in the front may
result from the shocks or from
annihilation and/or reconnection of
oppositely directed magnetic
fields in the front (Romanova and Lovelace 1992
and Lovelace, Newman, and Romanova 1996).

Section 2 derives a complete set of
equations for the front, specifically,
equations for the total number of particles, the
total momentum, energy, and magnetic flux in the
front. Section 3 discusses and summarizes the
behavior of the solutions of the front equations.
Section 4 gives comparisons of predictions of
the theory with observations.
Conclusions of this work are given in Section 5.

\begin{figure*}[t]
\begin{center}
\includegraphics[width=3.5in]{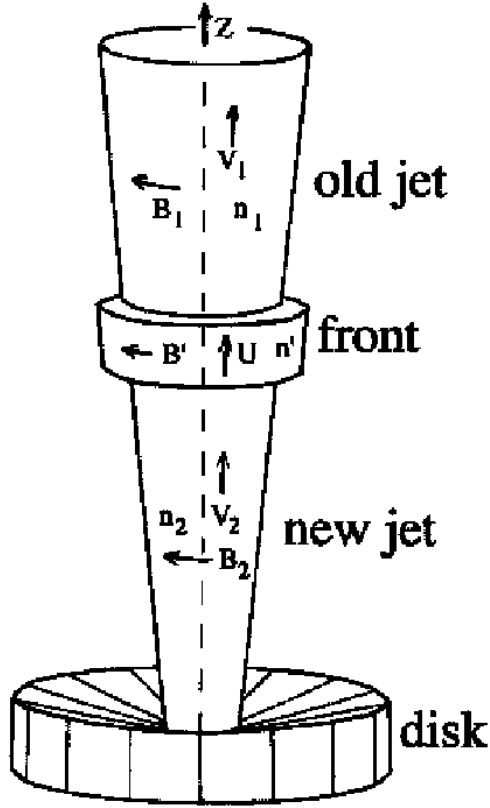}
\end{center}
\caption{Sketch of the geometry of a
propagating front in a Poynting flux jet.}
\end{figure*}

\section{THEORY}

We use an inertial, cylindrical
coordinate system $(r,\phi,z)$ with the
origin at the black hole's center
and the $z-$axis normal to the
accretion disk as shown in Figure 1. This is
referred to as the `lab frame'.
We consider that Poynting flux
jets propagate symmetrically away
from the disk in the $\pm z$ directions, but
focus our attention on the approaching
$+z$ jet. A Poynting flux jet is
{\it self-collimated}, with energy,
momentum, and angular momentum
transported mainly by the electromagnetic
fields (Lovelace, Wang, and Sulkanen 1987).
The collimation results from the toroidal
magnetic field at the edge of the jet.
A steady Poynting jet can be characterized
in the lab frame by its asymptotic
$(z\gg r_o)$ magnetic field
$B_\phi = - B[r_o/r_j(z)]$, and electric
field $E_r =- (v/c)B[r_o/r_j(z)]$ at the
jet's edge, $r=r_j(z)$, where $r_o$ is
the jet's radius at $z=0$, $v = const.$
is the jet's axial velocity, and $B$ is
the lab frame field strength at $z=0$.

The jet plasma consists of both ions
and leptons (electrons and positrons)
with the ratio of leptons to ions
$f_{li}$ The initial jet radius is
taken to be $r_o = 6GM/c^2$, where $M$ is
the black hole mass. The energy
flux (luminosity) of the $+z$ jet
is the Poynting flux $L_{j} = v B^2 r_o^2/8 =$ constant.

We propose that the Poynting
jet from the disk changes abruptly at
time $t=0$. That is, the jet parameters
change from values with subscript (1) to subscript (2) at
$t=0$, $ (~B_1,~n_{i1},~ v_1,~ f_{li1}) \rightarrow
(~B_2,~n_{i2},~v_2,~f_{li2})~ .$ In
actuality the change will be
with a time scale determined by the disk
dynamics as in the implosive accretion
model of Lovelace et al. (1994).  This
time scale may be as short as $r_o/c$ in
the lab frame.  In the
present work we consider $ b \equiv
{B_1 / B_2}<1, ~\nu \equiv {n_{i1}/
n_{i2}} , ~ v_1 < v_2,~ f_{li1} = f_{li2}, $
where $n_i$ is the number density of
ions. The change in the jet parameters
produces a `front' which propagates
outward as indicated in Figure 1. The
front may involve a pair of shock waves, one
for the incoming old jet matter and
the other for the incoming new jet matter.
We let $Z(t)=z(t)/r_o$ denote the
dimensionless axial distance of the front.
We also use lab time
$T = t{\big/}(r_o/c)$,
speed of the front
$U(T)= dZ/dT=(dz/dt){\big/}c$, and
bulk Lorentz factor $\Gamma = (1-U^2)^{-{1\over2}}$. We let
$V_{1,2}=v_{1,2}/c$ and
$\Gamma_{1,2}=(1-V_{1,2}^2)^{-{1\over 2}}$.
The time measured in the front
frame is $T' = t'{\big/}(r_o/c)$ with
$dT' = dT{\big/}\Gamma(T)$. The initial
values at $T=T'=0$ are $Z=0$ and $U=V_1$.
We also use the time
$T'' =t''{\big/}(r_o/c)$ measured by
a distant observer oriented at an angle
$\theta_{obs} < \pi/2$ to the $z-$axis;
$dT'' = dT [1-U~cos(\theta_{obs})]$. Further,
the observed time $T'''$ for a cosmological
source is given by $T''' = T''(1+z)$ with $z$ the redshift.

\subsection {Number of Particles}

The continuity equation for ions in the
front frame is $$ {{\partial n_i'}\over{\partial t'}} =
-{{\partial (n_i'v')}\over {\partial z'}}~,
\eqno(1) $$ where $n_i'$, $v'$, and $t'$
are the number density, velocity, and
time in the front frame. We integrate this equation over a
cylindrical 'pill box' of radius
$ r > r_j(z)$ and axial length
$\Delta z'$ and use the Lorentz transformations
to obtain an equation
for the total number of ions in the front $N_i(T)$,
$$
{{dN_i}\over{dT}} = N_{io}
\big[ \Delta V_2 H(\Delta V_2) +
\nu \Delta V_1 H(\Delta V_1) \big].
\eqno(2)
$$
Here, $N_{io} \equiv \pi r^2 n_{i2}(z) r_o
= \pi r_o^3 (n_{i2})_{z=0} $ is a
constant by our earlier assumptions.
Also, $H(x)=1$ for $x>0$ and
$H(x)=0$ for $x<0$. If $\Delta z'$ were a constant,
then we would
have $\Delta V_1 = U-V_1$ and
$\Delta V_2 = V_2-U$, where upper case
quantities are dimensionless. However,
the plasma in the front expands
freely with speed $C_s'$ (normalized by $c$)
so that $\Delta Z' = \Delta Z_o' + 2\int_0^{T'}
dT'' C_s'(T'')$, and this results
in modified expressions for $\Delta V_{1,2}$.
In the radial direction the
plasma also expands freely with sound speed
$C_s'$. $C_s'$ is the compressional wave
speed in the front frame. For the cases
studied, $C_s'$ is determined essentially
by the particles in that the particle
energy density is larger than that of
the magnetic field. If both the ions and
leptons are highly relativistic in the
front frame, $C_s'=1/\sqrt3$, whereas for
non-relativistic ions but
relativistic leptons $C_s' < 1/\sqrt3$.
Magnetic pinching of
the front is estimated to be small.
Note that the number density of ions
in the front frame is simply
$n_i' = N_i{\big/}(\pi r_o^3 R^2 \Delta Z')$. The electric field in
the front frame $|{\bf E}'|$ is small compared with $|{\bf B}'|$.

If $f_{li1}=f_{li2}$ and there is no
$e\pm$ pair production in the front,
then the total number of leptons in the front is
simply $N_l= f_{li}N_i$. However, in
the general case considered here,
$$
{{dN_l}\over{dT}} = N_{io}
\big[ f_{li2}\Delta V_2 H_2 +
f_{li1}\nu \Delta V_1 H_1 \big] +
{1\over\Gamma}{\bigg({{\delta N_l}
\over{\delta T'}}{\bigg)_{e^\pm}}}~, \eqno(3) $$
where the $H's$ are the same as in equation (2).
The main contribution to the pair production
for the conditions considered is from
collisions of synchrotron and SSC photons,
$$\bigg({{\delta N_l}
\over{\delta T'}}\bigg)_{e^\pm} =
(\pi r_o^3 R^2\Delta Z')~r_o \times$$ $$
\int d\epsilon_1 \int d\epsilon_2 ~
{{dn'_{syn}}\over{d\epsilon_1}}~{{dn'_{SSC}}
\over{d\epsilon_2}} \sigma_{pair}
\bigg|_{\epsilon_1\epsilon_2 >(m_ec^2)^2} .$$
\noindent For rough estimates, we can write
$\epsilon_1={3\over2} \gamma^2 \hbar
\omega_o'$, where $\omega_o' = e| B'|/(m_ec)$ is
the cyclotron frequency in the front frame,
and $\epsilon_2 = \gamma^2 \epsilon_1$.
Thus, a rough condition for an electron
to give pair production is
$ \gamma \geq \gamma_{pair}
\equiv \big(mc^2/\hbar\omega_o'\big)^{1 \over 3}$
$\approx 3.5\times 10^3 (10^3G/|B'|)^{1\over3}$.
Electron-positron recombination is negligible
for the conditions considered.

\subsection{Momentum Conservation}

In the front frame,
$$
{{\partial T_{0z}'}\over {\partial t'}}=
-{{\partial T_{zz}'}\over{\partial z'}} + grav + rad~, \eqno(4) $$

\noindent where $T_{0z}'$
and $T_{zz}'$ are components of the
energy momentum flux density tensor
in the front frame, and 'grav' and 'rad' denote
gravitational and radiative force contributions
not included in $T_{ij}'$. We integrate (4)
over the same pill box to obtain $$
\big(N_i m_i {\bar \gamma}_i + N_l m_e
{\bar \gamma}_l\big)~ \Gamma^3 {{dU}\over{dT}}
= - {{\pi r^2 r_o} \over{c^2}}\big [T_{zz}'\big] +$$
$${r_o\over c^2}\int (grav+rad)~. \eqno(5)
$$
${\bar \gamma}_i~$ and $~{\bar \gamma}_l~$
denote averages over the ion and lepton
distribution functions in the {\it front frame}.
In this frame the distributions are
assumed isotropic and that for electrons is
assumed the same as for positrons.
The initial values of ${\bar\gamma}_i$ and ${\bar
\gamma}_l$ are considered to be close to unity.
A Lorentz transform gives $
\big(T_{zz}'\big)_s = \Gamma^2\big(T_{zz}-2UT_{oz}+ U^2T_{oo}\big)_s, $
where $s=1,2$. The lab frame components of the energy-momentum
tensor for a Poynting flux jet are $$  T_{oo}=
T_{zz}={ {E_r^2 + B^2_\phi}\over {8\pi} }
= (1 + V^2)_s
\bigg({ {B^2} \over {8\pi} }\bigg)_s~,$$
$$ T_{oz}
= {{E_r B_\phi}\over {4\pi}}=2( V )_s
\bigg({ {B^2} \over {8\pi} }\bigg)_s~,  \eqno(6)
$$
where $B=B_\phi$ at the edge of the jet. Thus, in equation (5) we have
$$
-{{\pi r^2 r_o}\over{c^2}}\big[T_{zz}\big] =
{{r^2r_o B_2^2\Gamma^2}\over{8
c^2}}\bigg\{\big[(1+U^2)(1+V_2^2)-4UV_2\big] - $$
$$b^2\big[(1+U^2)(1+V_1^2)-4UV_1\big]\bigg\}~,
\eqno(7) $$

\noindent where $b^2 \equiv (B_1/B_2)^2$, and $r^2B_2^2 =
const =r_o^2(B_2)^2_{z=0}$. We assume $b^2 < 1$.
The terms in the square brackets
involving $V_2$ represent the push from the new
Poynting jet, while those involving $V_1$ are
for the push in the opposite direction from the old Poynting jet.
A small modification of equation (7) similar
to that of (2) is required to account for the
free expansion of the plasma of the front.
Note that the electromagnetic field
contribution to the momentum of the
front is small because $|{\bf E}'|\ll |{\bf B}'|$.

Dividing equation (5) by
$~\Delta M_o \equiv N_i(0)m_i+N_l(0)m_e~$ gives
$$ \bigg({{{N}_im_i{\bar \gamma}_i+
{N}_lm_e{\bar \gamma}_l}\over{\Delta M_o}} \bigg) \Gamma^3{{dU}\over{dT}} =
\mu \Gamma^2 \bigg \{\big[(..)\big]\bigg\} +$$
$$ {r_o\over{\Delta M_o c^2}}\int(grav + rad)~, \eqno(8) $$
where
${\mu} \equiv{ { r_o^3(B_2)_{z=0}^2}/({8 \Delta M_o c^2}})~\gg 1$
is a dimensionless measure of the strength of the Poynting jet.
The brackets
$\{\}$ denote the same quantity as in equation (7).  The
driving term $\propto \mu$ vanishes if both $\Gamma_1 \rightarrow
\Gamma_2$ and $b\rightarrow 1$.
If $\Gamma, \Gamma_1,$ and $\Gamma_2$ are all much
larger than unity, it is clear from equations (7)
and (8) that a steady state is possible if
$b^2 > (\Gamma_1/\Gamma_2)^4$ (recall that $b^2 <1$); that
is, the push of the old and new jets
balance and the Lorentz factor of the front is
$\Gamma =\Gamma_1 [(1-b^2)/(b^2 - (\Gamma_1/\Gamma_2)^4)]^{1\over 4}$.
If $b^2$ is smaller than $(\Gamma_1/\Gamma_2)^4$,
then no balance is possible and the Lorentz factor
of the front $\Gamma$ increases without limit.

The gravitational
force in equation (8) can be written as
$\int grav = -GM$ $(N_im_i+N_lm_e)/ (r_o^2+z^2)~. $
The radiative force depends in general
on the geometry and energy distribution
of the background radiation field of
the central region of the AGN.
Dermer et al. (1992) consider the case
where the radiation comes from the disk, while
Sikora et al. (1994) argue that the
radiation field is from disk radiation
scattered by a distribution of clouds
orbiting the central object. We adopt
a rough parameterization of the radiation
fields of Dermer et al. and Sikora et al. The average
photon energy is denoted
${\bar \epsilon}_{ph}$, the
total luminosity $L_{ph}$, and the
characteristic radius of the spatial
distribution $r_{ph}$. The force due
to this radiation field is $\int rad~
=~ { (N_l\sigma_T) L_{ph} {\cal D}_{pz}{\cal F}_T}
/ [{\pi (r_{ph}^2+(z-a r_{ph})^2)c]}$, where
$a=0$ for the Dermer et al.
model and $a=1$ for that of Sikora et al.
Here, ${\cal D}_{pz}=\Gamma^2[cos(\theta_{ph})-U]| cos(\theta_{ph})-U|$ is the
Doppler factor which accounts for the change in the flux and the
change in the $z$ momentum of the photons between the lab
and front frames, $cos(\theta_{ph})=
(z-ar_{ph})(r_{ph}^2+(z-ar_{ph})^2)^{-{1\over2}}$. $\sigma_T$ denotes the
Thomson cross section. The contribution of
leptons with $\gamma > m_e c^2/{\bar \epsilon_{ph}}$ to 
$(N_l\sigma_T)$ in the radiative
force is reduced in that the Klein-Nishina
cross section applies.
${\cal F}_T = (1-e^{-\tau_T})/{\tau_T}$ with $\tau_T=n_l'~r_o~R~\sigma_T$ the
Thomson optical depth of the front. For
the conditions considered
here, $\tau_T \ll 1$.
For small axial distances, $cos(\theta_{ph}) < U$,
the radiative force acts as a drag whereas at
larger distances it gives a push.

\subsection{Energy Conservation}

In the front frame,
$$
{{\partial T_{oo}'}\over {\partial t'}}
= - {{\partial T_{oz}'}\over {\partial z'}}
  -syn - ssc - Com~, \eqno(9) $$
\noindent where the last
three terms represent the energy loss
rates due to synchrotron radiation,
inverse-Compton scattering off of synchrotron
photons, and inverse-Compton scattering
off of the above mentioned background
photons. Following our
previous method, we integrate (9) over
the volume of the front to obtain
$$
\Gamma{d \over{dT}}
\bigg[ N_i({\bar \gamma}_i'-1)m_i
+N_l({\bar \gamma}_l'-1)m_e + {{W_B'}\over{c^2}} \bigg] =$$
$$ -{{\pi r^2 r_o}\over{c^3}} \big[ T_{0z}' \big]
- {{r_o}\over{c^3}} \int(syn+ssc+Com)~,
\eqno(10)
$$
\noindent where
$W_B' = (r_o^3/{4c^2}) R^2 \Delta Z' (B')^2 $  is
the magnetic field energy in the front.
The magnetic field $B'$ is discussed below in
subsection 2.4.  The driving term $\propto \mu$
vanishes if both $\Gamma_1\rightarrow \Gamma_2$
and $b\rightarrow 1$.
Because the acceleration process(es)
in the front is not known, we make the
well-known supposition
(for example, Pacholczyk 1970)
that the kinetic energy in the ions
is a constant factor $k$ times that in the
leptons, $N_i({\bar \gamma}_i-1) m_i = k N_l({\bar \gamma}_l-1)m_e~.$
As a result, ${\bar \gamma}_i$ is
determined in terms of ${\bar \gamma}_l$.
Therefore, in the following $\gamma$ without a
subscript refers always to the lepton Lorentz factor.

A Lorentz transform gives $T'_{oz}
= \Gamma^2 \big[(1 + U^2) T_{oz} - U T_{oo} - U T_{zz}\big]~, $ so
that
$$
-{{\pi r^2 r_o}\over {c^3}}\bigg[T_{0z}'\bigg]
= {{r^2 r_o B_2^2 \Gamma^2}\over{4 c^2}} \bigg
\{ \big[(1+U^2)V_2-U(1+V_2^2)\big]H_2 $$
$$-b^2\big[(1+U^2)V_1-U(1+V_1^2) \big]H_1\bigg\}~. \eqno(11)
$$
Dividing equation (10) by $\Delta M_o$ gives $$ {{\Gamma 
}\over{\Delta M_o}}{d \over{dT}}
\bigg[N_l({\bar \gamma}_l-1)(1+k)m_e
+ {{W_B'}\over{c^2}} \bigg] =
2{\mu \Gamma^2}
\bigg \{\big[(..)\big] \bigg\}$$
$$ - {{r_o}\over{\Delta M_o c^3}} \int(syn+..)~, \eqno(12)
$$
\noindent where the
brackets $\{\}$ denote the same quantity as in equation (11).
The driving term $\propto \mu$ vanishes if both $\Gamma_1 \rightarrow
\Gamma_2$ and $b \rightarrow 1$.

The synchrotron energy
loss rate in the front frame is:
$$\int syn~ =~{{32 \pi}\over{9}}
r_e^2 N_l \overline{(\gamma)^2}c
\big( {{(B')^2}\over{8 \pi}} \big){\cal F}_{syn}, \eqno(13a) $$
where $r_e \equiv e^2/(m_e c^2)$ is the classical electron
radius, and ${\overline {(\gamma)^2}}=
\int_{\gamma_1}^\infty d\gamma
\gamma^2 f_l{\big/}N_l$, where $\gamma_1$ is the Lorentz factor
below which
synchrotron self-absorption becomes strong as discussed below. ${\cal F}_{syn}
=[1-exp(-\tau_{syn})]/\tau_{syn}$ is
synchrotron opacity factor with $\tau_{syn}$
the optical depth for synchrotron photons (Pacholczyk 1970).
The synchrotron-self-Compton energy loss rate is:
$$ \int ssc ~= ~{4\over3} (N_l\sigma_T) {\overline{(\gamma)^2}}
c \big({{64 N_l\overline{(\gamma)^2}
\sigma_T}\over{9 r_o^2 R \Delta Z'}} \big)
\big( {{(B')^2}\over{8 \pi}} \big) {\cal F}_{syn}~. \eqno(13b) $$
The energy loss
rate in the front frame due to Compton
scattering off of the background
photons is: $$ \int com ~ = ~ {{4}\over{3}}
(N_l\sigma_T) {\overline{(\gamma)^2}} c
\big({\cal U}'_{ph}\big) {\cal F}_T~, \eqno(13c)
$$
where ${\cal U}_{ph}'=
L_{UV} {\cal D}_{ph}^2/[\pi r_o^2 c (R_{ph}^2 +(Z-aR_{ph})^2)]$
is the background photon energy density in the front frame, and ${\cal D}_{ph}=
\Gamma [1-U cos(\theta_{ph})]$ is the Doppler
factor for photons in the front frame relative to the disk.

\subsection{Magnetic Flux Conservation}

The magnetic field
in front frame $B'$ is determined by taking into
account the influx of magnetic flux with the new jet matter (subscript 2)
and the old jet matter (subscript 1). We let $\Phi'= R\Delta Z'B'$ denote
the toroidal flux (in Gauss) in the front. Then $$
{{d\Phi'}\over{dT}} = B_o
\big[ \Delta V_2 H(\Delta V_2) + b \Delta V_1 H(\Delta V_1) \big], \eqno(14) $$
where $B_o \equiv (B_2)_{z=0}$ and the $H's$
are the same as in equation (2). If $b < 0$,
there is annihilation of magnetic flux in the
front (see Romanova and Lovelace 1992; and Lovelace, Newman, and 
Romanova 1996).

\subsection{Lepton Distribution}

The lepton distribution function suggested by
observations has a hard power law form in the
main energy containing range, say, $f_l = K_1/\gamma^2$
for $\gamma_1 \leq \gamma \leq \gamma_2$ with $1\ll \gamma_1\ll\gamma_2$.
For $\gamma\geq\gamma_2$, the distribution is steeper,
say, $f_l=K_2/\gamma^3$, and at even larger energies,
$\gamma \geq \gamma_3$, $f_l$ is even steeper.
For $\gamma < \gamma_1$, $f_l$ is assumed negligible (see Figure 2). 
Thus, $f_l$
is characterized mainly by $\gamma_1$ and $\gamma_2$.
We have ${\overline \gamma}
=\gamma_1 [1+ ln(\gamma_2/\gamma_1)],$ and ${\overline {\gamma^2}}=
{\cal C}\gamma_1 \gamma_2,$ with ${\cal C} = 1+ln(\gamma_3/\gamma_2)$.
The relevant value of $\gamma_1$ is the
Lorentz factor below which synchrotron
self-absorption becomes strong. Self-absorption as
treated for example by Pacholczyk (1970) then gives
$\gamma_1=\gamma_1(B', n_l', R)$. With $\gamma_1$ known, we have
$\gamma_2=\gamma_1 exp({\overline \gamma}/\gamma_1 -1)$.
The value of ${\cal C}$ depends only
weakly on $\gamma_3/\gamma_2$ and we
take $\gamma_3/\gamma_2={\overline \gamma}/\gamma_1$.

\begin{figure*}[t]
\plotone{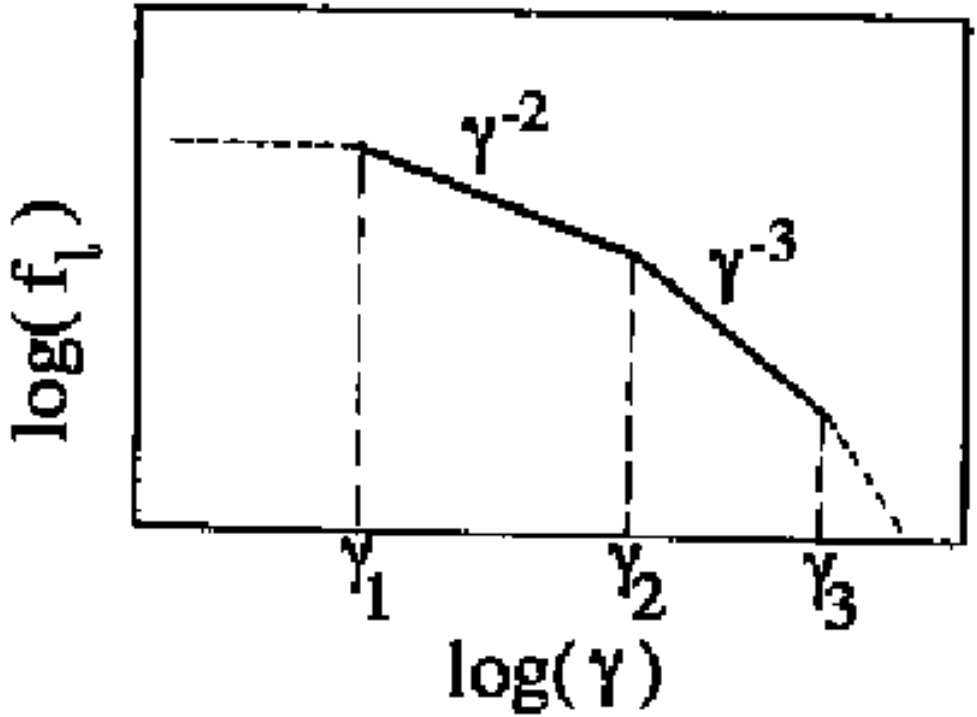}
\caption{Lepton distribution function in the comoving frame
of the front. $\gamma_1,~\gamma_2,$ and $\gamma_3$ are
derived from the equations of Section 2.}
\end{figure*}

\begin{figure*}[b]
\plotone{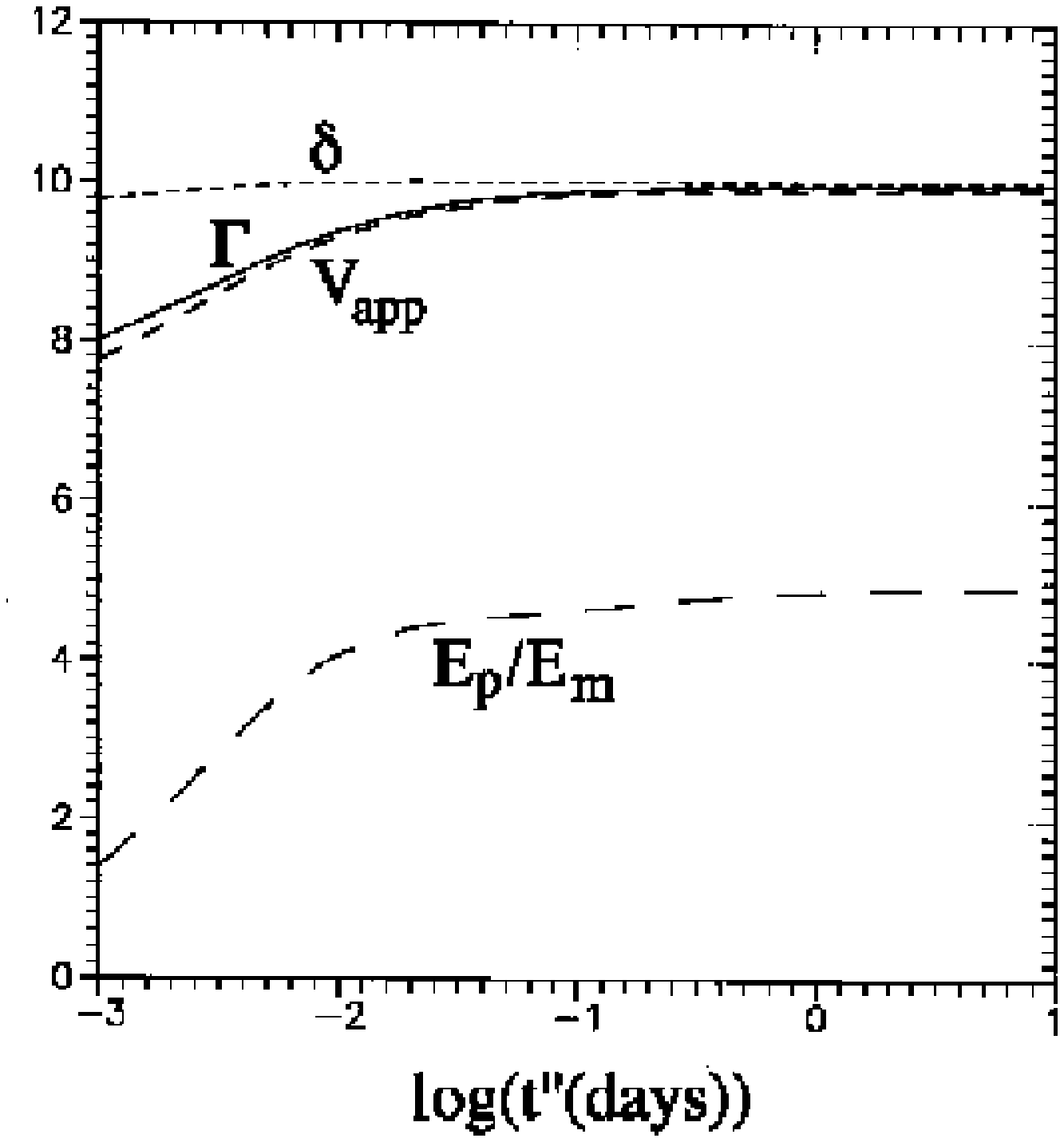}
\figcaption{Bulk Lorentz factor $\Gamma$, apparent
velocity of the front $v_{app}$ (normalized to c), and Doppler boost
factor $\delta$ versus time measured by a distant observer
for the reference parameters given in Section 3.}
\end{figure*}

\section{RESULTS}

Equations (2), (3), (8), (12) and (14) have been
solved numerically to obtain the time dependences
of the physical variables.
Here, we first discuss the time-dependences
for a reference case where
the black hole mass $M=10^9 M_{\odot}$ so that $r_o=8.9\times 10^{14}$ cm,
$\mu =r_o^3(B_2)^2_{z=0}/(8 \Delta M_o c^2)$ $=15$, $B_o=10^3$ G,
$b=|B_1/B_2| =0.5$, $\Gamma_1= 8$, $\Gamma_2=18$,
$\nu=n_{i1}/n_{i2} =0.44$, $k=1$, and the
initial lepton-ion ratio $f_{li}^o = 5$. For
the background photons,
$L_{UV}=10^{46}$ erg/s,
${\overline \epsilon}_{ph} = 10$ eV,
$r_{ph} = 0.1$ pc, and $a=0$.  The angle between
the observer and the jet axis is $\theta_{obs}=0.1$
rad. All results are presented for the
dependence on time ($t''$) seen by a
distant observer where it is recalled that
$dt''=dt [1-U cos(\theta_{obs})]$.
The redshift dependences are not
included in the equations.

\subsection {Time Dependences}

The velocity of the front
$U$, initially equal to $V_1$, increases
rapidly and approaches the equilibrium value
$U_{\infty}$ which corresponds to the
bulk Lorentz factor $\Gamma_{\infty} =10$.
This is shown in Figure 3.
The apparent velocity of the front
(with respect to the nucleus)
$V_{app} = U sin(\theta_{obs})
/[1-Ucos(\theta_{obs})]$ approaches $9.9$.
The Doppler boost factor $\delta
= 1/[\Gamma(1-U cos(\theta_{obs}))]$ approaches $10$.

\begin{figure*}[b]
\plotone{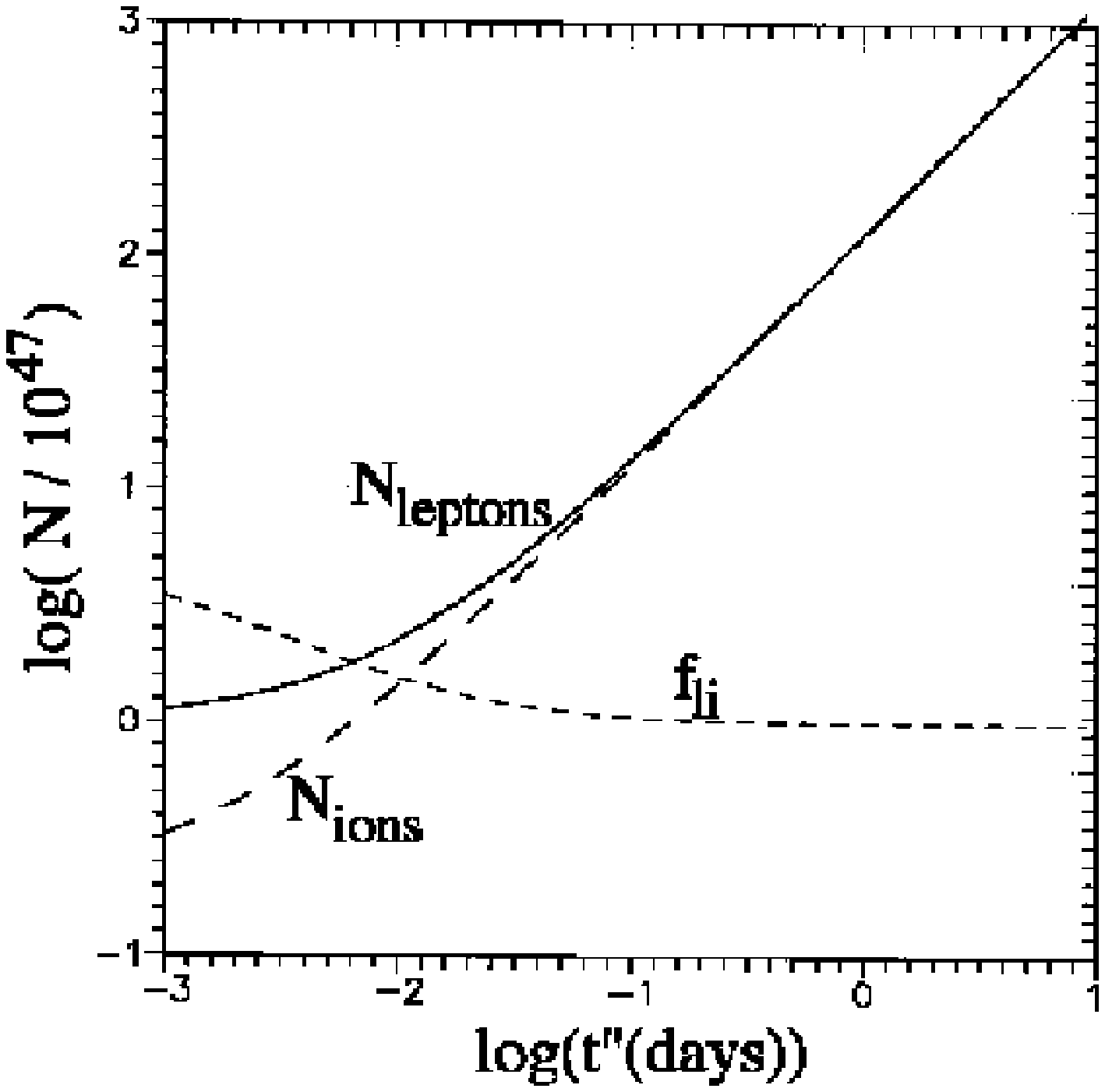}
\caption{Total number of leptons $N_l$ and ions $N_i$ in the front
and their ratio $f_{li}$ versus time measured by distant observer.}
\end{figure*}

The total number of
ions in the front grows with time,
because under most conditions the
front moves faster than the old jet
matter and slower than the new jet matter,
thus accumulating particles from both sides.
This is shown in Figure 4. For the case
shown there is no pair production. Thus, the ratio of leptons
to ions $f_{li}^o$ decreases with time
and approaches unity. Other cases were
calculated where pairs are created and the
number of leptons grows so that $f_{li}$ becomes
much larger than its intial value.

\begin{figure*}
\plotone{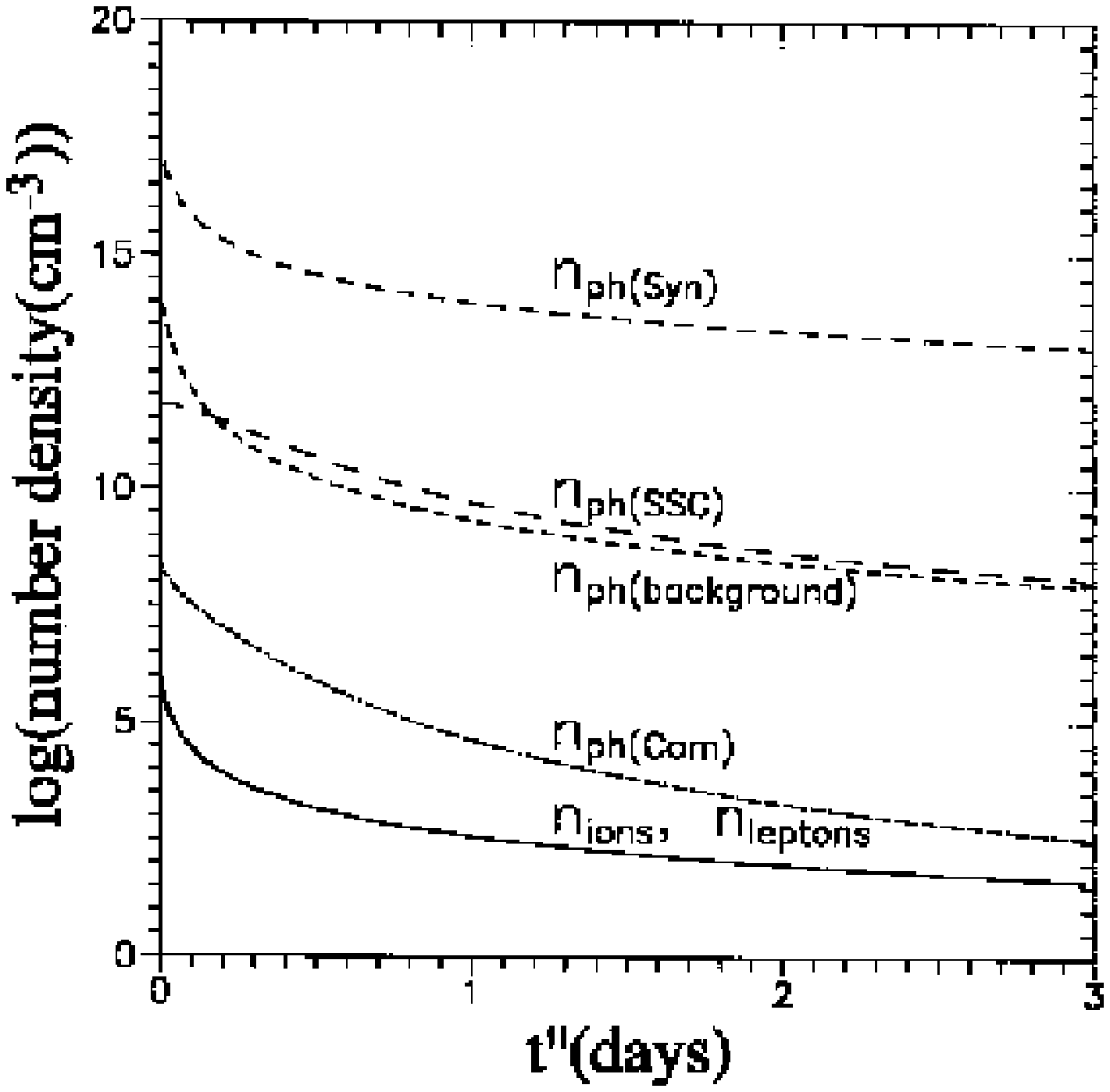}
\caption{Number density of leptons $n_l'$, ions $n_i'$,
background photons $n_{ph}'$, synchrotron photons $n_{syn}'$,
synchrotron-self-Compton photons $n_{SSC}'$ and
inverse-Compton photons $n_{Com}'$ (in units of $cm^{-3}$) in
the front frame versus the
logarithm of time measured by distant observer.}
\end{figure*}

Although the total number of particles
in the front grows with time, the
number densities decrease with time due
to the expansion of the volume of the
front. This is shown in Figure 5.
The number densities of synchrotron,
SSC, background, and inverse-Compton photons
in the front also decrease with time as shown in Figure 5.

\begin{figure*}
\plotone{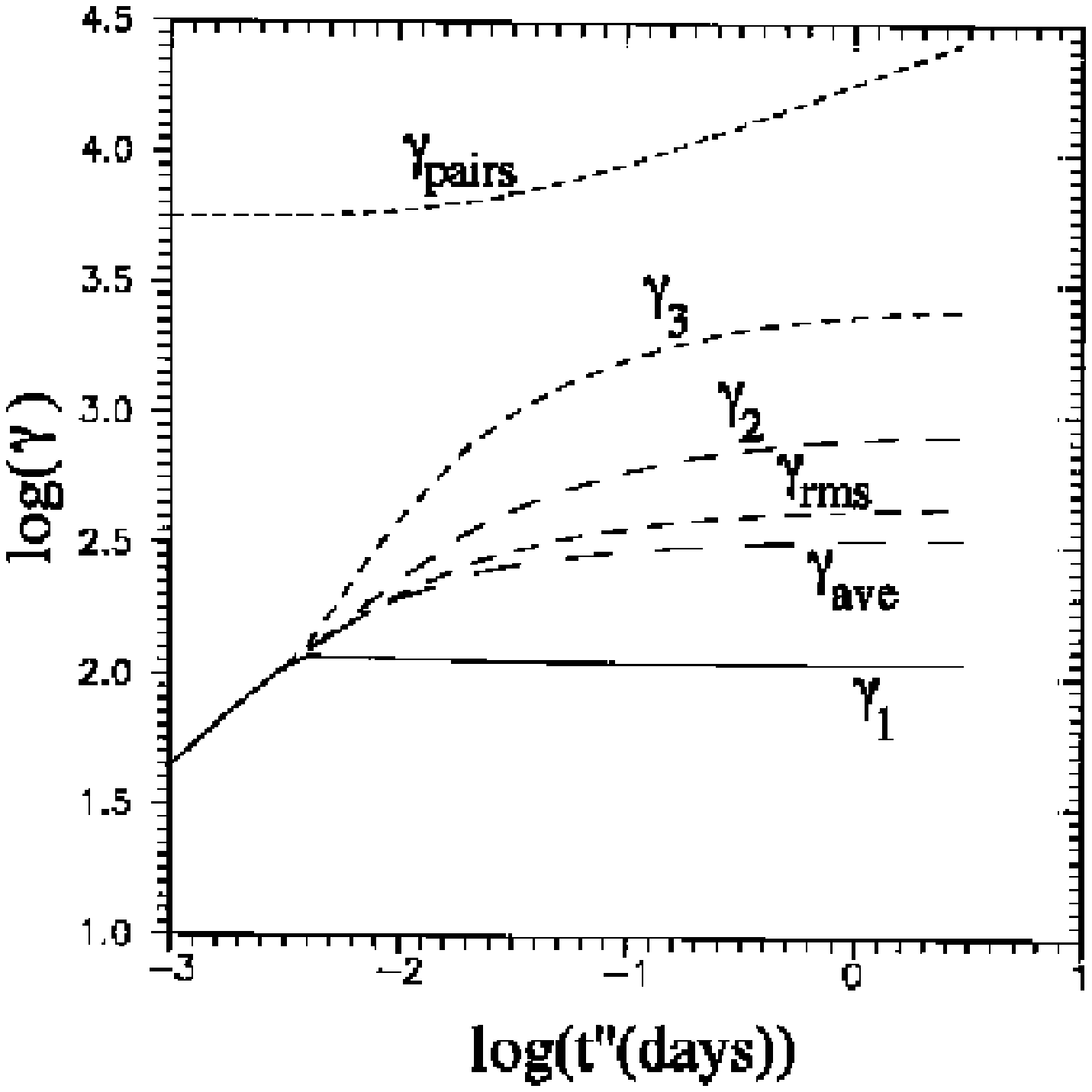}
\caption{Lorentz factors in the front frame separating
different parts of the lepton spectrum:
$\gamma_1, \gamma_2$, and $\gamma_3$
versus time measured by distant observer.
${\overline\gamma}$ is the average Lorentz factor.
$\gamma_{rms}$ is the the root mean square Lorentz factor.
$\gamma_{pairs}$ is the threshold Lorentz factor for pair
production. For the case shown there is no pair production.}
\end{figure*}


Figure 6
shows the evolution of the different
characteristic Lorentz factors, $\gamma_1,
\gamma_2,$ and $\gamma_3$, of the
lepton distribution. Initially, the
leptons are assumed to have the same
small energy corresponding to a
Lorentz factor $\gamma = 1.01$. Later,
when the average lepton Lorentz factor
($\overline \gamma$) becomes larger
than that corresponding to self-absorption
($\gamma_1$), a complete spectrum forms
extending from $\gamma_1$ to $\gamma_3$.
Initially, $\gamma_1, \gamma_2,$ and
$\gamma_3$ grow rapidly
and later more slowly. The top curve in
Figure 6, $\gamma_{pair}$, corresponds to the
energy of leptons at which pair creation
sets in.

The total apparent luminosities due to
synchrotron, SSC, and Compton processes
for a distant observer at an angle $\theta_{obs}$ to
the line of sight are calculated
including the solid angle boost factor $\delta^2$
between the front frame and the
observer, where $\delta=[\Gamma (1-U cos(\theta_{obs}))]^{-1}$.
These are shown in Figure 7.
At early times in an outburst, the SSC
radiation predominates because the
magnetic field in the front is strong
and the number of synchrotron photons
is higher than that of background
photons. Later, the magnetic field in
the front decreases and thus the
background photons predominate.
At the same time, the leptons are
accelerated in the front and
consequently inverse-Compton
radiation grows, and then later decreases.
The time of the peak of the inverse-Compton emission
depends approximately linearly on the
radius $r_{ph}$ of the background
photon distribution. Synchrotron
radiation is at much lower luminosity
during the first part of the front propagation.
The dependences of the maximum luminosities
of the different processes
are discussed later.

\begin{figure*}[t]
\plotone{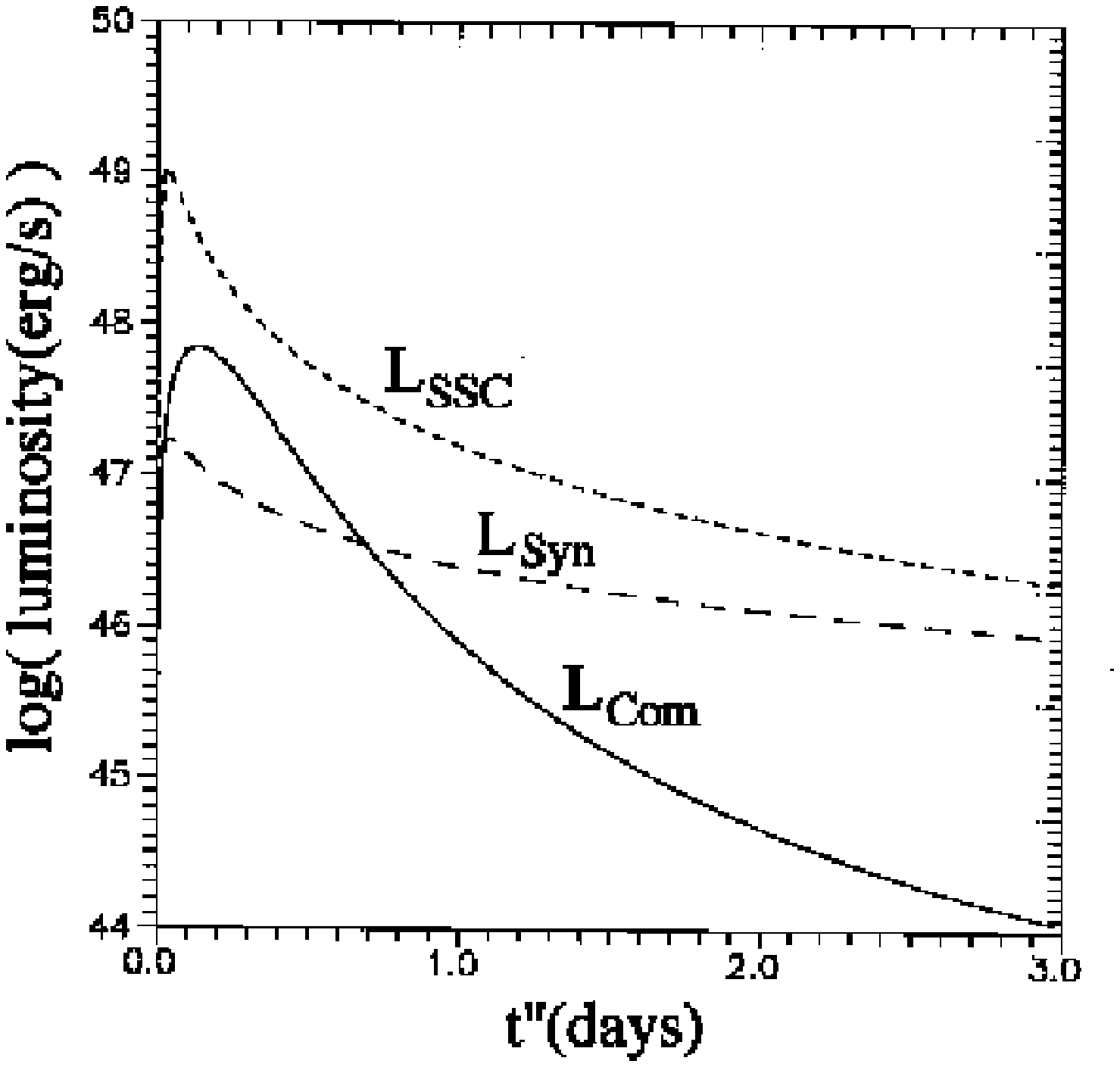}
\caption{Total (bolometric) apparent luminosities
radiated by the front due to the inverse-Compton, SSC,
and synchrotron processes (in $erg/s$)
as a function of time measured by distant observer.}
\end{figure*}


The observed frequency is boosted by a
factor $\delta$. The frequencies of radiation
corresponding to different processes
and to the characteristic values
$\gamma_1$, $\gamma_2$, and $\gamma_3$ of the
lepton distribution, are shown in
Figure 8. One can see that the radiation of
the front covers most of the frequency
range from the self-absorption limit
(the lowest curve), up to high gamma
ray energies. In many bands of the
spectrum, there are contributions
from two processes. For example, in
the Egret band (from $10^{22.5}$ Hz to $10^{24.5}$ Hz)
we typically have contributions from
both inverse-Compton and SSC radiation.
Initially, synchrotron radiation
contributes to the band from
IR to UV, SSC covers the range of
frequencies from soft X-ray to EGRET GeV
energies, and the inverse-Compton covers
the range from the high-energy X-ray to the EGRET GeV
energies. Thus, during an
EGRET high-energy flare one may expect
almost simultaneous
flashes in wavebands from IR to very high
energies, excluding the radio band
which is initially self-absorbed.

\begin{figure*}[b]
\plotone{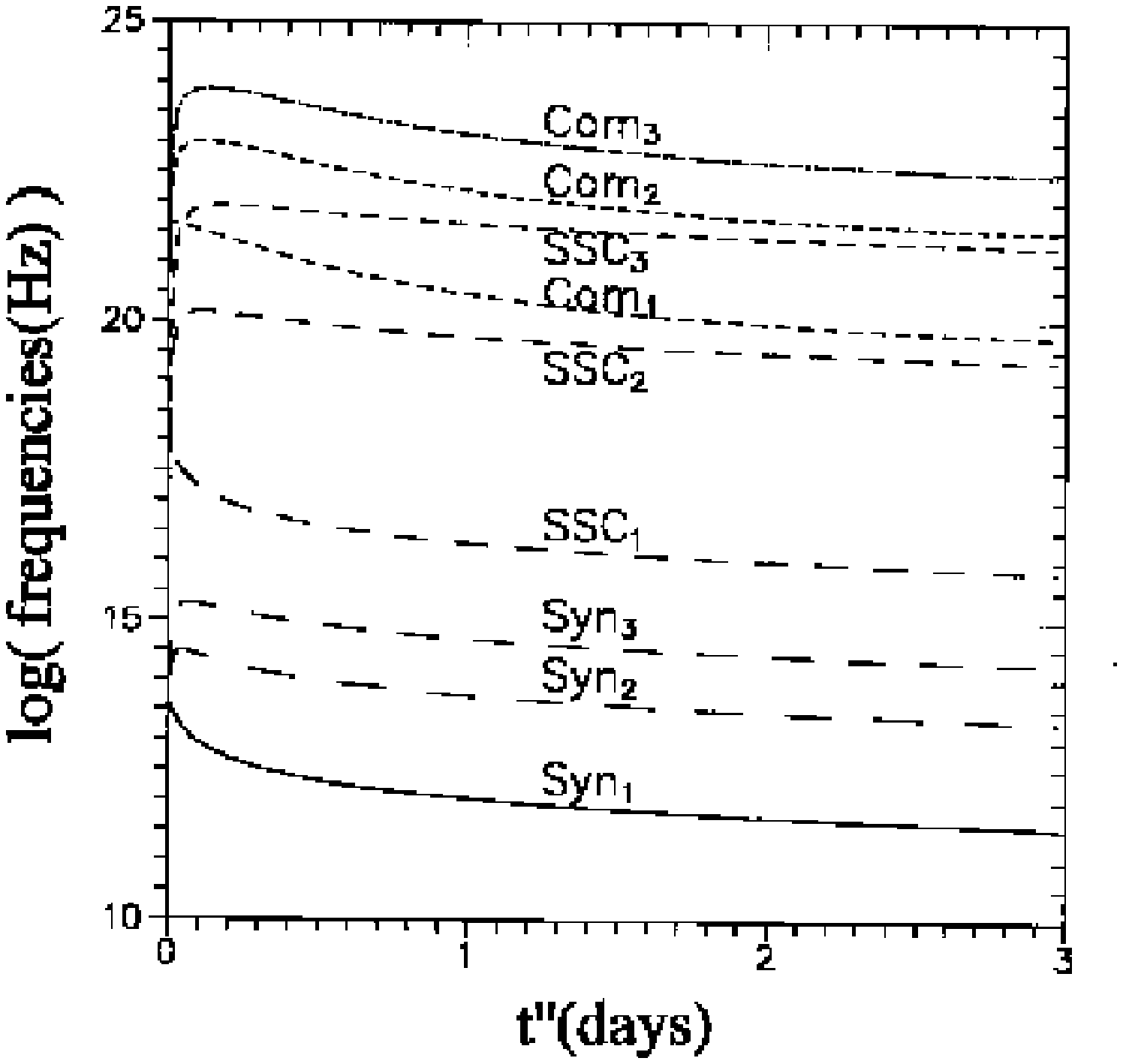}
\caption{Apparent frequencies in Hz of radiation
generated by electrons with Lorentz factors $\gamma_1,
\gamma_2$, and $\gamma_3$ due to the synchrotron, SSC,
and inverse-Compton processes as a
function of time measured by distant observer.}
\end{figure*}

\subsection{ Dependences of Maxima of Flares}

The maxima of the total apparent luminosities
due to inverse-Compton ($L_{Com}$),
synchrotron-self-Compton ($L_{SSC}$)
and synchrotron processes ($L_{syn}$)
depend on
the parameters
$\theta_{obs}$,
$B_o$, $\mu$, $M$, $f_{li}^o$,
$r_{ph}$, and $L_{UV}$ approximately as
$$
L_{Com} \sim 1.7\times 10^{47}
\bigg({0.1\over \theta_{obs}}\bigg)^{1.8}
\bigg({\mu\over 20}\bigg)^{1.4}
\bigg({M\over {10^9 M_\odot}}\bigg)^{1.0}\times$$
$$\bigg({B_o\over {10^3G}} \bigg)^{1.8}
\bigg({{10} \over{f_{li}^o}}\bigg)^{1.4}
\bigg({{0.1pc}\over {r_{ph}}}\bigg)^{1.0}
\bigg({L_{UV}\over {10^{46} ergs/s}}\bigg)^{1.0}~ {ergs\over s}~, $$ $$
L_{SSC} \sim 2.4\times 10^{47} \bigg({0.1\over \theta_{obs}}\bigg)^{1.8}
\bigg({\mu\over 20}\bigg)^{2.4}
\bigg({M\over {10^9 M_\odot}}\bigg)^{3.1}\times$$
$$ \bigg({B_o \over{10^3G}}\bigg)^{5.7}
\bigg({10\over {f_{li}^o}}\bigg)^{2.5}~{ergs\over s}~, $$
$$
L_{syn} \sim 1.6\times 10^{46} \bigg({0.1
\over \theta_{obs}}\bigg)^{1.8} \bigg({\mu\over 20}\bigg)^{1.3}
\bigg({M\over {10^9 M_\odot}}\bigg)^{2.2}\times$$
$$ \bigg({B_o \over {10^3G}} \bigg)^{3.8}
\bigg({10\over {f_{li}^o}}\bigg)^{1.3}~{ergs\over s}~, \eqno(15) $$
if the parameters are of the order of the normalization values.

The dependences on $\mu$ in equation (15) are approximately valid for
$10 \le \mu \le 50$.
For $\mu < 10$, the power is higher for SSC and
synchrotron radiation, but the same for Compton radiation.
For $\mu > 50$, the luminosities go to a constant.
The dependences on $B_o$ are valid for $B_o < 2000$ G,
and become flatter for stronger fields.
The dependence on $f_{li}^o$ is about the same as in
equation (15) for $5 < f_{li}^o < 15$, and becomes
steeper for SSC and synchrotron radiation for $f_{li}^o > 15$
and does not depend on $f_{li}^o$ for $f_{li}^o < 4$. As for
the dependences on $\theta_{obs}$, the $L's$
depend more strongly on $\theta_{obs}$
than in equation (15) at larger angles
($\theta_{obs}>0.1$) and more weakly for smaller angles. The
dependence on $M$ is approximately
the same at all parameter ranges.

Notice that the shape of the light
curves are quite different for
different parameters. There are
several possibilities: ({\bf1}) Both
SSC and Compton radiation may be
in the EGRET band. Then, the
shape of the EGRET light curve is
determined essentially by the
SSC radiation which gives a
fast rise of luminosity with typical time scale
of growth $t'' = 0.03$ days $= 43$ min
(see Figure 9a).  The rise time may of course be slower due
to a more gradual transistion from the old jet to the new.
({\bf 2}) SSC radiation
frequencies may be lower than the
EGRET band so that only Compton
radiation is observed. Then, the
light curve shape will be determined
by Compton radiation with a maximum
at $t'' = 0.13$ days for our reference values (see Figure 9b).
({\bf 3}) No radiation may fall in the
EGRET band. For example, changing $\mu$ (with other
parameters fixed) we get a sharp SSC
flare for $\mu > 35$, a flatter Compton
flare for $20 < \mu < 35$, and no radiation in
the EGRET band for $\mu < 20$. Thus, for
this set of parameters, the value $\mu$,
which is the initial ratio of magnetic
to rest mass energy-density, should be
relatively high. The shape of the EGRET
light curves also depends, for example,
on the initial ratio of leptons
to ions. For $1 < f_{li}^o < 5$,
SSC type flares predominate. For $5 < f_{li}^o < 15$,
Compton type flares predominate,
and for $f_{li}^o > 15$, no radiation
falls in the EGRET band. Thus, the
shape of the flare may be different
in different sources or may change
in the same source between different
outbursts, depending on the magnetic/particle
rest mass energy-density ratio, or on the initial ratio
of leptons to ions in the front
plasma.

\begin{figure*}[b]
\begin{center}
\includegraphics[width=3.5in]{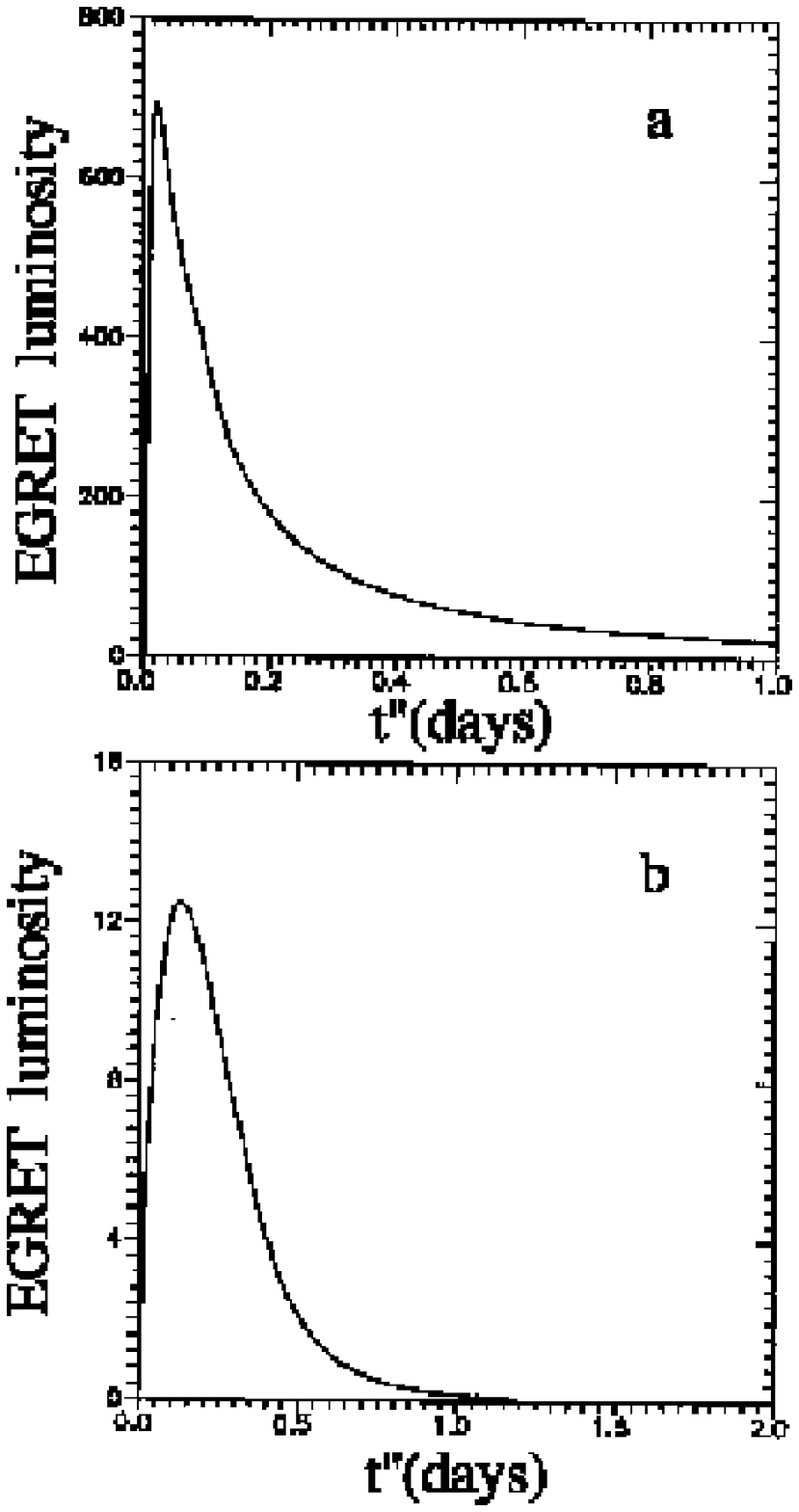}
\end{center}
\figcaption{Typical light curves of apparent luminosity in
the EGRET band in units of $10^{47}$ erg/s
as measured by distant observer.
In case (a), $f_{li}^o=10$,
and the light curve is determined
by SSC processes. In case (b), $f_{li}^o=20$,
and the light curve is
determined by the inverse-Compton process.}
\end{figure*}


The maxima of the flares in the SSC and
synchrotron radiation occur very
soon after outburst (much less than a day).
The maximum of the inverse-Compton flare is much later
and may be observable,
$$
t_{Com}''=0.13 \bigg({\theta_{obs}\over 0.1}\bigg)^{0.9}
\bigg({10\over \mu}\bigg)^{0.1}
\bigg({f_{li}^o \over 10}\bigg)^{0.1}
\bigg({r_{ph}\over {0.1 pc}}\bigg)~, \eqno(16) $$
in days.
Note that $t_{Com}''$ has essentially no dependence on $M$.

Later, after the maxima, all luminosities decrease. The SSC
luminosity decreases as $L_{SSC}\sim (t'')^{-2}$, the
synchrotron luminosity as $L_{syn}\sim (t'')^{-1}$,
and the inverse-Compton luminosity as
$L_{Com}\sim (t'')^{-4}$ for $t'' < 2-3$ days, but
later for $t > 15$ days it flattens to $\sim (t'')^{-1}$ .

The rise of the inverse-Compton luminosity is
determined by the growth of the number and
energy of leptons in the front. The front is
transparent to inverse-Compton
radiation from the beginning. At the same time,
initially plasma of the front is opaque to
the synchrotron photons, and the rapid
growth of synchrotron and SSC radiation is
determined by the fact that plasma
becomes transparent to synchrotron photons.
The decline of inverse-Compton
radiation is due to the fact that the energy density
of background photons
decreases in the front frame.
An important factor in the decrease
is the Doppler
shift of the background photons
relative to the front for $Z>R_{ph}$.
As for synchrotron and SSC radiation,
they decrease with time because the
magnetic field in the front decreases with time.

\section{COMPARISON WITH OBSERVATIONS}

\subsection{Gamma-Ray Variability:
Comparison with the Flare in PKS 1622-297}

Here, we compare theoretical and
observed light curves.
The luminosity in the
EGRET band $(10^{22.5}-10^{24.5})$ Hz is
determined in general by both inverse-Compton and
SSC processes. We integrated
the gamma radiation in the EGRET band
and got two qualitatively different types of light curves
(Figures 9a and 9b).
The difference is determined by the
fact that in some cases (depending on
parameters) SSC radiation is below the
EGRET frequency limit and the light curve
is determined by the inverse-Compton
light curve (see Figure 9b). In the opposite
case, when the SSC radiation is within
the EGRET band, the light curve is
determined by
both processes, but the SSC radiation
has a strong peak at small time scales,
and this determines the shape of the
light curve (Figure 9a). In the
first case, it is more probable
to observe the fast
rise and gradual decrease of gamma
luminosity, compared with the second
case, where the rise of luminosity is so fast
(maximum reached at $t''=0.02$
days) that it will be difficult to observe and only the `tail' of the
flare will be seen (see, however, Mattox et al. 1996).

EGRET observations of gamma-ray blazars
show different types of flares. In some
cases the growth of the luminosity is relatively
slow compared with the decay
(Kniffen et al. 1993). In other cases
(for example, Hartman et al. 1993),
only a decaying light curve is
observed during the EGRET set of
observations.
In the third type of
flare, the rise of the luminosity is
much faster than the decay (Mattox et al. 1996).

Time resolved
observations of the brightest observed
gamma-ray flare from blazar PKS 1622-29
(Mattox et al. 1996) show
that during the main flare the flux
increases by an order
of magnitude on a 2 hour time scale,
whereas the decay is on a 7 hour time scale.
Thus the rise is faster than the decay.
We fitted this flare by a `Compton type' flare using the
parameters: $M =
3\times 10^9 M_\odot$, $B_o = 1700$ G,
$\theta_{obs} = 0.1$ rad $= 11.5^o$, $\mu = 20$,
$f_{li}^o = 6$, $r_{ph}=0.3$ pc, ${\overline\epsilon}_{ph}=10$ eV,
$L_{UV}=2 \times10^{46}$ ergs/s, and $a=1$.
This is shown in Figure 10.
For calculation of the theoretical flux we used standard
formulae for the luminosity distance at the redshift
of the source $z=0.815$, Hubble's constant $H_o= 75$ km/s/Mpc,
and a cosmological parameter $q_o=1/2$.  Also, we took
into account the cosmological redshift of the frequencies and
the dilation of the time scale of the outburst.  The value
$a=1$ of Figure 10 corresponds to background photons scattered
by clouds (Sikora et al. 1994), and it gives a more
symmetric Compton flare than $a=0$.  We also obtained
fits to the PKS flare with $a=0$ where the background
photons come from the disk (Dermer et al. 1992).

The fit of our model to the PKS 1622-29
data is not unique, but the
fitting in all cases requires a relatively
massive black hole ($M > 10^9 M_\odot$)
and a relatively strong magnetic
field at the base of the jet ($B_o > 10^3 G$).
Also, it appears necessary to
have an initial ratio
of magnetic field to rest mass energy-density large compared with
unity, $\mu \sim 10 - 30$.
Note that the quasar PKS 1622-297 is the brightest
gamma-ray quasar (Mattox, et al., 1996).
In other lower gamma luminosity blazars,
$M$ and/or $B_o$ may be smaller.

The appearance of one type of
flare or another, and also the strength
of the flares, depends on
values of the plasma
parameters ($f_{li}^o$, $\mu$, $B_o$) which
may be random.  Flares where the stage of
growth is very short may be explained
by `spike' type SSC flares (Figure 9a),
whereas the slower rising gamma light
curves may be inverse-Compton flares
(Figure 9a). Note, that the fast growth
and decay of luminosity
in the beginning of the strong PKS 1622-29 flare may be
connected with `spike' type SSC flare, which appears at
smaller initial lepton-ion number ratios ($f_{li}^o < 5$).

In the present model, we proposed that the new Poynting
jet pushes matter continuously. However, if the
push decreases with time, then the
front will lose energy as a result of
radiation and the luminosity will decrease
more rapidly than that show here.
This effect may also explain
rapid decay, compared with growth, of the
prominent flare of the quasar 3C 279 (Kniffen et al. 1993).

\subsection{ Radio Variability}

Analysis of the correlation between
gamma-ray flares and those in radio
emissions at short time scales has shown that there is no
direct correlation in most blazars
(Reich et al. 1993).
Moreover, often the radio flux is in the
low state during the gamma-ray flare. However,
observations on longer time-scales show that
after strong flares in the EGRET band, the radio
luminosity starts to grow and has a maximum
several months later than the gamma-ray flare
(Reich et al. 1993). This may result
from either
synchrotron self-absorption
or free-free absorption by
plasma surrounding the source core
(Matveyenko et al. 1992). Thus,
the radio emission appears at
later times when the density
in the front and/or the external density are much lower.

In our model the synchrotron
self-absorption frequency is initially
in the infrared or UV, depending
on parameters. Thus, initially, there
is no radio emission from
the front. Later, the self-absorption
frequency $\nu_1(syn)$ slowly decreases
(Figure 8) and goes to $\sim 3$ mm
($10^{11}$ Hz) after several days.
Much later, $\nu_1(syn)$ falls into the cm radio band.
For some parameters of the model, the
millimeter emission appears
one or two weeks after the flare.
Initially, only a small fraction of the
total luminosity
goes into low frequencies, and the radio
flux is very small.
Because of the slow decrease of
$\nu_1(syn)$ with time,
$\nu_1(syn) \sim 10^{12} (t'')^{-1}$ Hz,
the
maximum of the radio luminosity may be a
few months later than the gamma-ray flare.
Here, we did not take into account possible
free-free absorption of radio waves by
distributed plasma in the core of the source
exterior to the jet.
This absorption may also contribute to the
delay in the appearance of the radio emission.

The present model of a propagating front in a Poynting flux jet
is similar in some
respects to the model of
radio variability of Aller, Aller, and
Hughes (1985) and Hughes, Aller, and Aller (1985) based on
weak shocks
propagating along a hydrodynamic jet.
Detailed multi-frequency radio
observations of the quasar 0528+134
shows that after a strong gamma-ray
flare, the radio flux grows
and has a maximum few months later
(Pohl et al. 1995). It is important to
note
that during this event, the
new VLBI super-luminal component
appeared with a back-extrapolat-ed
time of origin close to the time of the
gamma-ray flare (Krichbaum et al, 1995).
These observations support the present
model where a strong impulsive
disturbance - the front - propagates
along and is powered by an otherwise
invisible Poynting flux jet. The front
radiates first in gamma rays, and later
becomes strong in the radio when it is
observed as a new super-luminal VLBI jet component.

\begin{figure*}[b]
\plotone{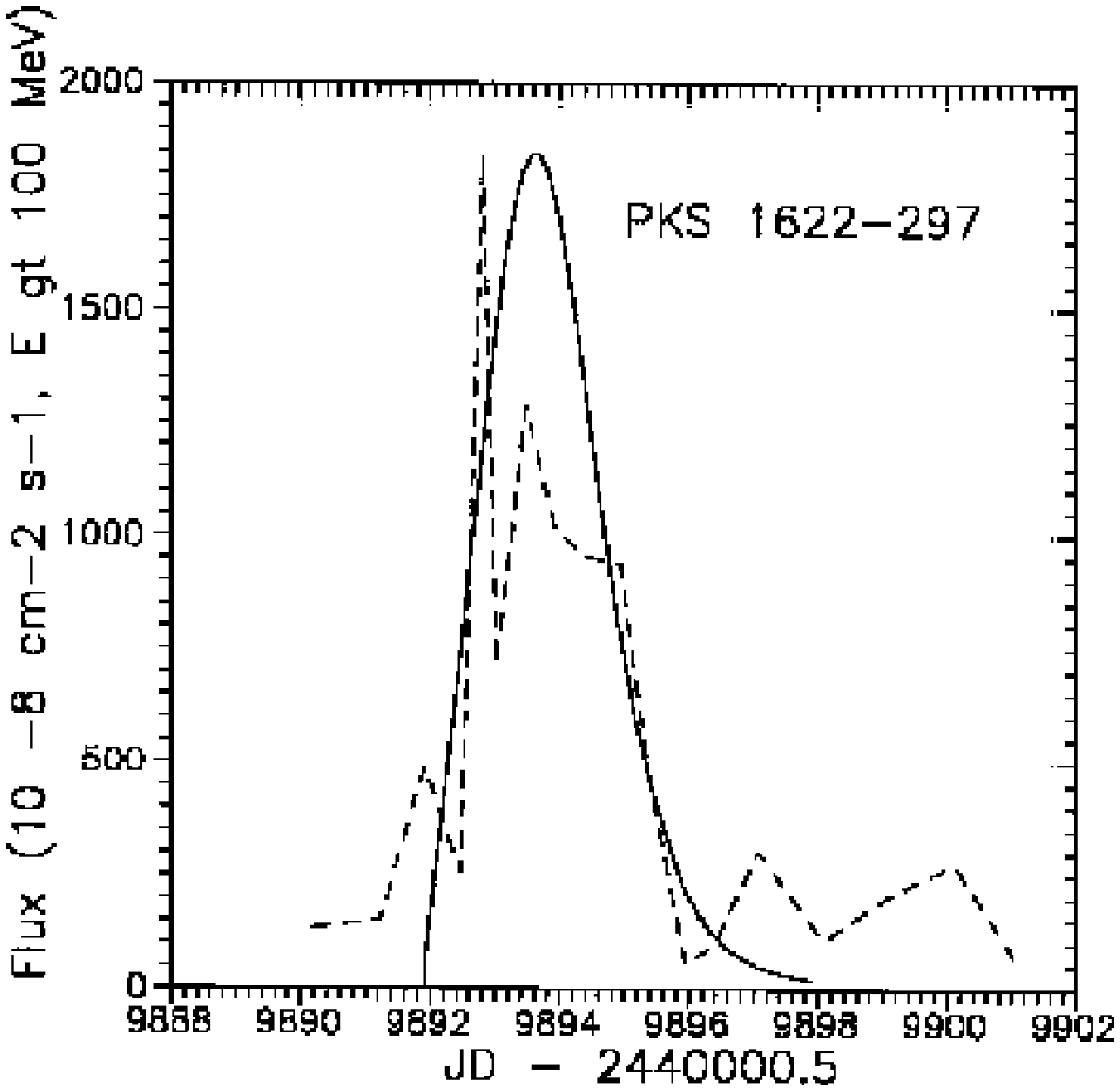}
\caption{Comparison of the EGRET light curve of
PKS 1622-297 (the dashed curve) (Mattox et al. 1996)
with the calculated `Compton type' light curve (solid curve).}
\end{figure*}


\section{CONCLUSIONS}

The paper develops a self-consistent dynamical
model of gamma-ray flares and VLBI outbursts of
blazars based on lepton
acceleration in the propagating front
of an otherwise invisible Poynting flux jet.
Inverse-Compton scattering of the accelerated, relativistic
leptons off of background photons as
well as off of synchrotron photons (the SSC process) and
synchrotron radiation are taken into account.

It is shown that gamma-ray flares in
the frequency band of the EGRET instrument
(at $50 < E < 10^4$ MeV) may be of
two types: ({\bf 1.}) `SSC spikes'
which have a fast rise ($\sim$ hr.) and gradual
decay of luminosity. This type of
flare appears in the cases where SSC
photons have high enough energy to be
in the EGRET band. In this case inverse-Compton
radiation, which usually gives
a smaller contribution to the luminosity, is hidden
by the strong SSC flare. ({\bf 2.}) `Compton flares'
which have a growth time of few hours and a
duration of days up to weeks (depending on the
distribution of background photons and viewing angle).
Comparison with recent
observations of the gamma-ray
flare in PKS 1622-297 (Mattox et al. 1996) show
that it may be interpreted as a `Compton flare'. To explain
the high luminosity and the short duration of
the PKS flare, the viewing angle must be quite small,
$\theta_{obs} \sim 0.1$ rad.  Also, we need a black hole
mass $M \sim 3 \times 10^9 M_\odot$ and a
magnetic field at the base of the jet $B_o \sim 10^3$ G.

The model gives almost simultaneous
flares for photon energies from high-energy
gamma-ray to lower energy gamma, X-ray, UV,
visible, and down to the infrared. The short wavelength
(mm) radio flare for some parameters may
appear shortly ($\sim$ days) after the main gamma flare, because
the self-absorption frequency, which
initially corresponds to the IR region
of spectrum, decreases rapidly during a
few days (see Figure 8) to short wavelength
radio band (in some cases, however, mm
radiation may appear much later).
Later, it decreases more slowly, so that
the centimeter waveband radio radiation in
  most of the cases should appear much
later ($\sim$ months) than the gamma-ray flare.

In many gamma ray loud objects, both flares and
steady radiation are observed ( Fichtel and Thompson 1994;
Montigny et al. 1995).
The steady
radiation may come from a superposition of
a number of overlapping outbursts.  In other
cases, the EGRET instrument detects only the
highest states of strong flares with the weaker
steady emission obscured in the noise.

Extremely high energy photons ($\sim$ TeV)
have been detected from the object
Markarian 421 (Punch et al. 1992).
Here, we comment on the possible
origin of TeV photons in
the framework of the present model.
TeV radiation may result from
from the SSC process by lepton scattering off of
synchrotron photons if the
lepton distribution function extends to
sufficiently high Lorentz factors
($\sim 10^5$). However, the magnetic
field at the base of the jet $B_o$
may have to be smaller than the values
considered here ($\sim 10^3$ G) in
order to avoid excessive pair production.

\acknowledgments{
Both authors were
supported in part by NSF grant AST-9320068.
MMR was also supported in part by
Russian Fundamental
Research Foundation Grant No 93-02-17106
and by the Scientific and Educational Center
of Kosmomicrophysics "KOSMION".
We thank Drs.
J.R. Mattox, T.P. Krichbaum, and
A. Levinson for stimulating discussions.}

\end{document}